# Bose-Einstein condensation of excitons in planar systems, and superconductive phase transition temperature

Y.Ben-Aryeh

*Physics Department Technion-Israel Institute of Technology, Haifa, 32000, Israel*

*E-mail:* phr65yb@physics.technion.ac.il   (Y.Ben-Aryeh)

**Abstract**

A theoretical model is developed for treating super conductive Bose-Einstein condensation (BEC) effects for excitons in planar systems, under the condition that many excitons are included in a surface area, with the dimensions of the exciton center of mass de Broglie (dB) wave length, and under the condition that attractive forces are introduced between different excitons. The total internal energy of the excitonic system is found to be composed of the separate excitons positive energies, and negative energy due to scattering between different excitons. We assume that for high density of excitons, and corresponding attractive interactions between excitons, excitons in internal mode $k$ are annihilated and in the same time excitons in internal mode $k'$ are created, where these scattering effects are integrated for all $k$ and $k'$ values. It is assumed that the internal momenta of the excitonic system, remains in a quasi-stationary state, with approximately Bose distribution. Self-consistent equation for BEC of electrons-holes pairs and corresponding phase transition temperature are developed as function of electromagnetic interactions, experimental conditions and parameters. Possible applications by the use of the present theory are described.

1. Introduction

Quite long ago it has been suggested that quasi-particles, such as 'excitons', i.e., non-localized states of solids, can fulfill necessary conditions for a Bose-Einstein condensation (BEC).[1] Analyses of BEC of excitons [2] in $Cu^2O$, in potential-traps,[3] and for exciton-polaritons,[4] have been reported. BEC excitonic effects have been described in some review articles .[5–7] A book about BEC of excitons and bi-excitons has been published.[8] The possibilities for BEC of excitons have been discussed.[9]

A.A. High et al.[10], described experiments for BEC in gas of indirect excitons, confined in an electrostatic trap. By following his interpretation, we find that the boson nature of excitons allows for condensates at low temperatures, but because of recombination excitons have a finite lifetime that is too short to allow cooling to low temperature in regular semiconductors. The exciton lifetime should be large compared to the exciton cooling time and this requirement can be fulfilled in gas of indirect excitons. An indirect exciton is a bound state of an electron and hole in separable layers. The special structures for indirect excitons[11–13] enable the control of various parameters for the excitonic systems. For high density of excitons in planar, or semi planar, systems the observation of BEC of excitons can be related to coherent luminescence from all the excitons included in a coherent area.[1–14]

As is well known condensates are the basis of superconductivity based on electron-electron–Cooper-pairs. In ordinary superconductors, two spin1/2 electrons bind into a Cooper pair which is a boson. These bosons can undergo BEC in which a macroscopic number of Cooper pairs enter into a single wave function. In principle the same thing should be possible with pairs states made of electrons and holes, where the attraction between the fermions is not phonon mediated effect, but due to Coulomb attraction between particles with positive and negative charge. In the present work, we develop a certain theoretical model which can lead to superconductive Bose condensation in bilayer systems,[15] which are similar those developed for indirect excitons.[10–13]

Previous theories for exciton BEC are involved with many controversies. Superconductive excitons effects have been predicted by certain theories,[16–22] which have been developed quite long ago, but experiments have not confirmed, so far, such predictions. It has been claimed in certain works[23–24] that excitons do not have perfectly bosonic commutation relations at high densities. We consider, however, excitons in quite low temperatures where the bosonic commutation relations for excitons are expected to be valid, at least approximately. The coherence properties of indirect excitons have been studied,[25–28] showing long coherence times, thus improving the conditions for superconductivity effects. On the other hand it has been claimed that quantum signature in indirect excitons is blurred by disorder.[29] It has also shown in a certain analysis,[30] that there is a crossover between the mean-field and the fluctuating regime, at a certain phase transition point. In some studies,[31] it has been found that in strong electro-magnetic (EM) fields the excitonic system is turned into a plasma state. We assume, however, that the strength of the EM-fields, for the system which will be treated in the present work, is below such critical phase transition conditions.

There are many articles analyzing exciton-exciton interactions.[32–33] Especially, such analyses are studying possible binding energies for bi-excitons.[34–36] It has been claimed[33] that in addition to the one component fluid phase, an excitonic fluid phase, a bi-excitonic fluid phase can be produced for very small bilayers separations, and the transition between these phases appears to be continuous. M.A.M. Versteegh et al.[37], reported an observation of electron-hole Cooper pairs in highly excited $ZnO$. I find, however, that the binding energies are too small, for producing many bi-excitons. I analyze in the present work, as follows, a much stronger effect:

For high density of excitons attractive forces are produced between different excitons leading to annihilation of excitons, at internal mode $k$, and in the same time excitons at internal mode $k'$ are created and such scattering effects are made over all values of $k$, $k'$. The total internal energy of the present excitonic system is composed of the separate excitons energy plus scattering negative energy, due to attractive forces between different excitons. Under certain conditions the system can enter into a quasi-stationary state for which Bose Einstein distribution can be applied, at least approximately. As we will use a mean-field theory the scattering effects will be taken into account by an average negative potential. The present study shows that under certain conditions, which are analyzed in the present work, the system can enter into a superconductive state with a band gap $\Delta$. It is shown that superconductive, BEC of electrons-holes Cooper pairs can occur in excitonic planar systems, under the condition that many excitons are included in a surface area, with the dimensions of the exciton center of mass de Broglie (dB) wave length (often referred as the BCS limit), and under the condition that a negative potential is produced due to scattering effects between different excitons. Self-consistent equation for Bose condensation of electrons-holes and corresponding phase transition temperature will be developed as function of electromagnetic interactions, experimental conditions and parameters. The present theory is different from BCS theory as it is based on electromagnetic interactions, and a "semi-divergence" of the exciton distribution in planar systems, depending on a certain parameter $\alpha$, which is different from that in bulk media.

The present paper is arranged as follows:

In Section 2, I analyze the Boson electrons-holes-pairs distribution in planar systems. In Section 3, I develop a quantum mean field for scattering effects leading to BEC, with energy gap $\Delta$. The quantization is made for a large number of excitons included in an area, with the exciton center of mass dB wave length dimensions, taking into account their internal momenta distributions, as excitons localizations and their spatial fluctuations might not be described in x, y space, within this surface area. In Section 4, I summarize the present results and discuss possible applications of the present theory.

## 2. Electrons-holes-pairs distribution in planar systems

The transition of an electron from the filled valence band to the conduction band[38] obtained by the absorption of a photon, is producing an electron-hole pair with energy

$$\varepsilon(k) = \frac{\hbar^2 k^2}{2}\left(\frac{1}{m_{el}} + \frac{1}{m_{hole}}\right) + E_{gap} - E_{bind} = \frac{\hbar^2 k^2}{2m_{eq}} + E_{gap} - E_{bind} \quad , \tag{1}$$

where $m_{el}$ and $m_{hole}$ are the effective mass of the electron and hole in the conduction and valence band, respectively, $m_{eq}$ is the equivalent mass for the electron-hole pair, $E_{gap}$ is the band gap energy and the attraction energy between electron and hole can be taken into account by the negative energy $-E_{bind}$. We notice that the energy $\varepsilon(k)$ is given by a function of the momentum $k$, of the equivalent mass $m_{eq}$, the binding energy of the electron-hole pair $-E_{bind}$ and also of the energy band gap $E_{gap}$. While there might be different perturbations which will produce the electrons-holes pairs, the essential point is that annihilation and creation operators for such states satisfy boson commutation relations. We assume that electrons-holes are produced in the plane or semi-planar surfaces.

Let us assume that we have an ideal $2D$ gas of excitons with mass $m_{eq}$ that occupy area $A$. The total number $N$ of these particles in the area $A$ with temperature $T$ is given by:[39]

$$N = \int_0^\infty D(E) f(E) dE \qquad . \qquad (2)$$

$f(E)$, is the occupation number given by the Bose-Einstein distribution

$$f(E) = \frac{1}{\exp\{\beta[\varepsilon(k) - \mu]\} - 1} \qquad . \qquad (3)$$

Here $\mu$ is the chemical potential, $\varepsilon(k)$ is the single particle energy and $\beta = 1/k_B T$. The $2D$ density of states $D(E)$ can be given as:[39]

$$D(E) = A \frac{m_{eq}}{2\pi\hbar^2} \times \begin{cases} 1, & E \geq E_0 \\ 0, & E < E_0 \end{cases} \qquad . \qquad (4)$$

Here $E_0$ denotes the excitation energy with zero kinetic energy. For the simplicity of analysis we assume only one kind of excitons, but (4) can be generalized by assuming several kinds of excitons. We can change in (2) the integration over $E$ to integral over $k$ by using the relation

$$dE = d\varepsilon(k) = d\left(\frac{\hbar^2 k^2}{2m_{eq}}\right) = \frac{\hbar^2 k dk}{m_{eq}} \qquad . \qquad (5)$$

By substituting, (1) and (3-5), into (2) we get:

$$\langle N \rangle = A \frac{1}{(2\pi)^2} \int dk \frac{2\pi k}{F \exp\{\beta[(\hbar^2 k^2 / 2m_{eq}) + E_{gap} - E_{bind}]\} - 1} \quad ; \quad F = \exp(-\beta\mu) \, . \quad (6)$$

Eq. (6) can be interpreted as follows: $A$ represents the area in which excitons are located, and $2\pi k dk$ represents the $2D$ differential area for internal momentum of the excitons. Multiplying $A$ with $2\pi k dk$, we get the differential area in a certain phase space and by dividing it by $(2\pi)^2$ we get the differential number of modes. The total number of excitons is obtained in (6), by multiplying the differential number of modes by the occupation number, according to Bose statistics given by (3) and (1), and by integrating over k. One should take care, however, of the fact that the internal momentum $k$ is not representing any translational momentum of the exciton center of mass.

Substituting $k^2 = x$, $dx = 2kdk$ in (6), we get:

$$\begin{aligned} \frac{N}{A} &= \frac{1}{4\pi} \int dx \frac{1}{F \exp\{\beta[(\hbar^2 x / 2m_{eq}) + E_{gap} - E_{bind}]\} - 1} \\ &= \frac{1}{4\pi} \int dx \frac{F^{-1} \exp\{-\beta[(\hbar^2 x / 2m_{eq}) + E_{gap} - E_{bind}]\}}{1 - F^{-1} \exp\{-\beta[(\hbar^2 x / 2m_{eq}) + E_{gap} - E_{bind}]\}} \end{aligned} \quad . \quad (7)$$

The critical point in the present calculation is that we get explicit result for the surface density of electrons-holes pairs but the integration should be taken into certain limits as follows:

$$\frac{N}{A} = \frac{2m_{equ}}{4\pi\beta\hbar^2} \ln\left[1 - F^{-1} \exp\{-\beta[(\hbar^2 x / 2m_{eq}) + E_{gap} - E_{bind}]\}\right]_{x=0}^{x=k_f^2} \quad ; \quad F = \exp(-\beta\mu) \, . \quad (8))$$

Here we find that we have a certain band of electrons-holes pairs energies where in the lower limit we have $x = 0$, i.e., $\vec{k} = 0$ and in the upper limit $x = k_f^2$. Assuming that we have obtained experimentally a certain surface density $N/A$ of electrons-holes pairs, at temperature T, then the value of the chemical potential $\mu$ is a function of this surface density.

Under the condition that $\left[\left(\hbar^2 k_f^2/2m_{eq}\right)+E_{gap}-E_{bind}\right]\beta \gg 1$ the upper limit in (8) can be neglected and then we get

$$\frac{N}{A} = -\frac{m_{eq}}{2\pi\beta\hbar^2}\ln\left[1-F^{-1}\exp\{-\beta(E_{gap}-E_{bind})\}\right] \quad . \tag{9}$$

Eq. (9) can gives $N/A$ as function of $F$.

By using the anti-log in (9) we get:

$$\exp\left[-\frac{2\pi\beta\hbar^2}{m_{eq}}\frac{N}{A}\right] = \left[1-F^{-1}\exp\{-\beta(E_{gap}-E_{bind})\}\right] . \tag{10}$$

From (10) we get the relation

$$F\exp\{\beta(E_{gap}-E_{bind})\} = \frac{1}{1-\exp\left[-\dfrac{2\pi\beta\hbar^2}{m_{eq}}\dfrac{N}{A}\right]} \quad ; \quad F = \exp(-\beta\mu) . \tag{11}$$

Then, by substituting this result in (6) we get

$$N = \frac{A}{2\pi}\int\frac{kdk}{\exp\{\beta\hbar^2 k^2/2m_{eq}\}\left\{1-\exp\left[-\dfrac{2\pi\beta\hbar^2}{m_{eq}}\langle n\rangle\right]\right\}^{-1}-1} = \frac{A}{4\pi^2}\int n(k)2\pi kdk \quad . \tag{12}$$

Here $\langle n\rangle = \dfrac{N}{A}$ is the average number of excitons per unit area, and the number of excitons with internal momentum between $k$ and $k+dk$ is given by $\dfrac{An(k)k}{2\pi}$, where

$$n(k) = \frac{1}{\exp\{\beta\hbar^2 k^2/2m_{eq}\}\left\{1-\exp\left[-\dfrac{2\pi\beta\hbar^2}{m_{eq}}\langle n\rangle\right]\right\}^{-1}-1} \quad . \tag{13}$$

$n(k)$, can be considered as the density of excitons numbers in internal momentum space. The distribution of (13) depends on the correction term

$$\alpha = \exp\left[-\frac{2\pi\beta\hbar^2}{m_{eq}}\langle n \rangle\right] \quad . \tag{14}$$

In order to demonstrate the effect of this correction term, included in the squared brackets of (13), let us put some numbers. Assuming, for example: $m_{eq} = 0.1 m_{elec}$, $T = 10^0 K$ (where $m_{elec}$ is the usual electron mass), then

$$\exp\left[-\frac{2\pi\beta\hbar^2}{m_{eq}}\langle n \rangle\right] \approx \exp\left[-5.6\cdot 10^{-11}\langle n \rangle\right] \quad . \tag{15}$$

Here $A$ is given in units of $cm^2$. Under the condition $\langle n \rangle \gg 10^{11}$ the expression (15) becomes quite small and the term $(1-\alpha)^{-1} \simeq (1+\alpha)$ tends to value 1, but even under this condition we should take into account the effect of $\alpha$ as it eliminates the divergence of $n(k)$ in (13) for $k$ tending to zero.

There is a fundamental difference between cooperative effects in a planar system and those of a three-dimensional system. The average distance between $N$ boson particles in a three dimensional system is given by $(V/N)^{1/3}$, where $V$ is the volume, while for a planar systems it will become smaller as is given by $(A/N)^{1/2}$ where $A$ is the area of the system. So that attractive forces between different excitons might be more important for planar systems.

## 3. Electrons-holes-pairs with high surface density leading to BEC with energy gap

For high surface density of electrons-holes exciton, where many excitons are included in an area of the exciton center of mass dB wave length dimensions, scattering effects might be produced between different excitons. If $w_k$ is the probability that the exciton is produced, then the energy of such electron-hole-pair is given by

$$E_k = w_k \mathcal{E}_k = w_k \left( \frac{\hbar^2 k^2}{2m_{eq}} + E_{gap} - E_{bind} \right) . \tag{16}$$

Since electron-hole pair state can be either occupied or unoccupied, we choose a representation of two orthogonal states $|1\rangle_k$ and $|0\rangle_k$ where $|1\rangle_k$ is the state in which electron-hole excitonic state is occupied, and $|0\rangle_k$ is the corresponding unoccupied state. The general state of the electron-hole pair is given

$$|\psi\rangle_k = u_k |0\rangle_k + v_k |1\rangle_k \quad ; \quad w_k = v_k^2 \quad ; \quad 1 - w_k = u_k^2 . \tag{17}$$

Here for simplicity of calculations we assumed that the amplitudes $u_k$ and $v_k$ are real. The ground state of the many-body system can be described by the product

$$|\Psi\rangle = \prod_k \left( u_k |0\rangle_k + v_k |1\rangle_k \right) . \tag{18}$$

In this approximation the many-body state can be described as a superposition of occupied and un-occupied excitonic states. In two-dimensional representation

$$|1\rangle_k = \begin{pmatrix} 1 \\ 0 \end{pmatrix}_k \quad , \quad |0\rangle_k = \begin{pmatrix} 0 \\ 1 \end{pmatrix}_k . \tag{19}$$

The creation and annihilation operators for electrons-holes pairs are described, respectively, by

$$\sigma_k^+ = \begin{pmatrix} 0 & 1 \\ 0 & 0 \end{pmatrix} \quad ; \quad \sigma_k^- = \begin{pmatrix} 0 & 0 \\ 1 & 0 \end{pmatrix} . \tag{20}$$

The total Hamiltonian of scattering between different excitons, is given here by summation over all pairs interactions as:

$$H = -\sum_{k,k'} V_{k,k'} \sigma_k^+ \sigma_{k'}^- \quad . \tag{21}$$

Eq. (21) can be described as scattering effect where an exciton with wave vectors $k'$ is annihilated, and exciton with a wave vectors $k$ is created, due to a negative potential $V_{k,k'}$, where such scattering is summed for all $k$ and $k'$ values. These scattering effects for the internal momentum of the excitons can be related to the average attraction between two excitons, in the coherence area, by the planar Fourier transform:

$$V_{k,k'} = \int V(r) \exp[-i(k-k')r] 2\pi r dr \quad . \tag{22}$$

Here for simplicity we assumed that the external attractive potential $V(r)$ between two excitons is dependent only on the relative distance $r$ of the center of mass of these two excitons. This attractive energy between two excitons is coupled to the difference in internal momentum $(k-k')$ of the two excitons, which represents also a change in their internal energies by using (1). Thus (21) and (22), describe the coupling between external attractive forces between excitons and their internal momenta, or correspondingly internal energies.

The essential point in our analysis is that we treat excitonic systems with high surface density of excitons and low temperatures. Then the dB wavelength of the exciton center of mass is given by

$$\lambda_{dB} = \left( \frac{2\pi\hbar^2}{m_{eq} kT} \right)^{1/2} \quad . \tag{23}$$

We treat the present system under the condition that the dB wavelength is very large relative to the average distance between excitons. Assuming in an example: T=2K, $m_{eq} = 0.2 m_{elec}$ where $m_{elec}$ is the ordinary electron mass, then we get $\lambda_{dB} \approx 1400$ Angstrom. For surface density of excitons $n \approx 10^{12} cm^{-2}$ we get a corresponding average distance $d \approx 100$ Angstrom, so that approximately 200 excitons are included within a surface area of dB wave length dimensions. For higher surface density and/or lower temperatures this ratio is increasing further.

The present analysis is based on the idea that excitons center of masses are not localized in a region with dimensions smaller than the dB wavelength. Therefore when many excitons are included in a surface area with dB wavelength dimensions, we should treat the many body effects for internal momentum based on planar Bose statistics. The summation $-\sum_{k,k'} V_{k,k'} \sigma_k^+ \sigma_{k'}^-$ is given in a short notation for a summation over all internal modes $k$ and $k'$, where the change in their internal momentum, follows from potentials $V_{k,k'}$ related to attractive forces between ecxitons by the planar Fourier transform (22). We take into account that the distribution for $\langle n_k \rangle$ of (13) is very narrow, for $\alpha$ tending to zero and low temperatures. Then, under the conditions of the present analysis, we can neglect the dependence of $V_{k,k'}$ on $k - k'$ and exchange it into average value $V_0$. Then, the interaction Hamiltonian of (21) is given approximately by

$$\mathrm{H} \approx -V_0 \sum_{k,k'} \sigma_k^+ \sigma_{k'}^- \quad . \tag{24}$$

A straight-forward calculation shows that the total attractive energy is given by

$$\langle \Psi | \mathrm{H} | \Psi \rangle = -V_0 \sum_{k,k'} v_k u_k u_{k'} v_{k'} \quad . \tag{25}$$

Adding the energy of non-interacting pairs given by (16) to the attractive energy of (25), the total energy of the ground state is given by

$$W = \sum_k v_k^2 \varepsilon_{\bar{k}} - V_0 \sum_{k,k'} v_k u_k u_{k'} v_{k'} \quad . \tag{26}$$

We define

$$v_k = \cos\theta_k \quad ; \quad u_k = \sin\theta_k \quad . \tag{27}$$

Then (26) can be written as

$$W = \sum_k \varepsilon_k \cos^2\theta_k - \frac{1}{4}V_0 \sum_{k,k'} \sin(2\theta_k)\sin(2\theta_{k'}) \quad . \tag{28}$$

For the minimum of $W$ we get

$$\frac{\partial W}{\partial \theta_k} = -\varepsilon_k \sin(2\theta_k) - \frac{V_0}{2}\cos(2\theta_k)\sum_{k'}\sin(2\theta_{k'}) = 0 \quad . \tag{29}$$

From (29) we get:

$$-\varepsilon_k \tan(2\theta_k) = \frac{V_0}{2}\sum_k \sin(2\theta_k) \quad . \tag{30}$$

In (30) the summation over $k'$ has been exchanged into summation for $k$, since $k$ and $k'$ are equivalent variables. We define

$$\Delta = -\varepsilon_k \tan(2\theta_k) \quad ; \quad E_k = \sqrt{\varepsilon_k^2 + \Delta^2} \quad . \tag{31}$$

Then, we get:

$$E_k^2 = \varepsilon_k^2 + \Delta^2 = \Delta^2(1+\cot^2(2\theta_k)) \rightarrow \Delta = E_k \sin(2\theta_k) \quad . \tag{32}$$

We find

$$\sin(2\theta_k) = 2u_k v_k = \frac{\Delta}{E_k} \quad ; \quad \sum_k \sin(2\theta_k) = \sum_k \frac{\Delta}{E_k} \quad . \tag{33}$$

We substitute (33) into (30), using (31), then we get

$$\Delta = \frac{V_0}{2}\sum_k \frac{\Delta}{E_k} = \frac{V_0}{2}\sum_k \frac{\Delta}{\sqrt{\varepsilon_k^2 + \Delta^2}} \quad . \tag{34}$$

Consistency demands that $\Delta$ can be calculated by

$$1 = \frac{1}{2} V_0 \sum_k \frac{1}{\sqrt{\varepsilon_k^2 + \Delta^2}} \quad . \tag{35}$$

Following the above analysis $\Delta$ is the BEC energy gap.

The consistent relation given by (35) is similar to a consistent equation given by BCS,[40] but the distribution in internal momentum space, due to electromagnetic interactions, included in the summation of (35) is completely different, as will be analyzed here. It is different from a self-consistent equation obtained in previous works for excitons related to BCS theory,[16–22] since this equation has been obtained here only by assuming planar Bose statistics for high density of excitons where the attractive interaction between excitons is coupled with changes in excitons internal momenta, or correspondingly with their internal energies. As this equation can be applied only by inserting the full Bose statistics in the summation for $k$, the development of this equation gives results which are essentially different from those of BCS and from those obtained previously about the self-consistent equation for excitons.[16–22]

In order to take into account the summation over all excitons numbers which have wave-vector $k$, we replace the summation by integral using (12-13) as

$$\sum_k \to \frac{A}{(2\pi)^2} \int 2\pi k \langle n(k) \rangle dk \quad . \tag{36}$$

Here, A is the coherent area and for simplicity we assume that it equal to $\lambda_{dB}^2$, where $\lambda_{dB}$ is the de Broglie exciton center of mass wavelength. Then by using the relation (36) in (35), the self-consistent equation becomes

$$1 = \frac{V_0 \lambda_{dB}^2}{4\pi} \int k dk \{\varepsilon_k^2 + \Delta^2\}^{-1/2} \langle n(k) \rangle \quad . \tag{37}$$

Substituting in (37), $\varepsilon_k$ from (1) and $\langle n_k \rangle$ from (13), and changing the variables by $k^2 = x$ ; $dx = 2kdk$ then the self-consistent equation becomes

$$1 = \frac{V_0 \lambda_{dB}^2}{8\pi} \int dx \left[ \left( \hbar^2 x / 2m_{eq} + E_{gap} - E_{bind} \right)^2 + \Delta^2 \right]^{-1/2} \left\{ \frac{\exp\left( \beta \hbar^2 x / 2m_{eq} \right)}{1 - \alpha} - 1 \right\}^{-1} \quad . \tag{38}$$

We used here the definition (14) of $\alpha$:

$$\alpha = \exp\left[ -\frac{2\pi \beta \hbar^2}{m_{eq}} \langle n \rangle \right] \quad .$$

Eq. (38) can be simplified further by using the following change of variables:

$$\hbar^2 x / 2m_{eq} = E \quad ; \quad \beta \hbar^2 x / 2m_{eq} = \frac{E}{kT} \quad ; \quad dx = \frac{2m_{eq}}{\hbar^2} dE \quad . \tag{39}$$

Then (38) can be written as

$$1 = \frac{V_0 \lambda_{dB}^2}{4\pi} \frac{m_{eq}}{\hbar^2} \int dE \left[ \left( E + E_{gap} - E_{bind} \right)^2 + \Delta^2 \right]^{-1/2} \left\{ \frac{\exp(E/kT)}{1 - \alpha} - 1 \right\}^{-1} \quad . \tag{40}$$

Substituting $\lambda_{dB}$ from (23) into (40) we get

$$1 = \frac{V_0}{2kT} \int dE \left[ \left( E + E_{gap} - E_{bind} \right)^2 + \Delta^2 \right]^{-1/2} \left\{ \frac{\exp(E/kT)}{1 - \alpha} - 1 \right\}^{-1} \quad . \tag{41}$$

Eq. (41) represents the main theoretical result of the present paper. The transition temperature $T_{crit}$ for electrons-holes Cooper pairs BEC can be obtained from (41), under the assumption $\Delta = 0$. Then, for temperatures smaller than $T_{crit}$ the self-consistent equation (41), will be satisfied by inserting a certain value of $\Delta$ in this equation. However, if the right side of (41) will be smaller than 1 for $\Delta = 0$, then Bose condensation with energy gap $\Delta$ cannot occur. One should notice that the expression in the curled brackets represents Bose planar statistics effect and very small values of $\alpha$ ($\alpha \ll 1$) are needed for getting superconductivity effects. This conclusion is essentially different from that obtained in previous works.[16-22]

We find that the self-consistent equation with $\Delta = 0$ is given by

$$1 = \frac{V_0}{2kT}\int dE \frac{1}{(E+E_{gap}-E_{bind})}\left\{\frac{\exp(E/kT)}{1-\alpha}-1\right\}^{-1} \quad . \tag{42}$$

The integral in (42) is referred as a plane Bose integral and it is tabulated in the Appendix, as function of the parameters $(E_{gap}-E_{bind})/kT$ and $(1-\alpha)^{-1}$. $E_{gap}$, $E_{bind}$ and $\alpha$ have been defined in (1) and (14), respectively. This table demonstrates that this integral tends to have large values for small values of $\alpha$ ($\alpha \ll 1$).

It is convenient to measure the energies in units of $kT$ using the definitions:

$$\tilde{E} = E/kT \quad ; \quad \tilde{E}_{gap} = (E_{gap}-E_{bind})/kT \quad ; \quad \tilde{V}_0 = V_0/kT \quad . \tag{43}$$

Then the self-consistent equation with $\Delta = 0$ is given by:

$$1 = \frac{\tilde{V}_0}{2}\int d\tilde{E} \frac{1}{(\tilde{E}+\tilde{E}_{gap})}\left\{\frac{\exp(\tilde{E})}{1-\alpha}-1\right\}^{-1} \quad . \tag{44}$$

Equation (44) depends on three parameters .i.e., $\tilde{V}_0 = V_0/kT$, $\tilde{E} = (E_{gap}-E_{bind})/kT$ and $\alpha$. The conditions for superconductivity effects are improved for large values of $V_0/kT$ and smaller values of, $(E_{gap}-E_{bind})/kT$ and $\alpha$. In the present analysis we assumed that $\alpha$ is very small ($\alpha \ll 1$) corresponding to the condition of many excitons in surface area with the exciton center of mass dB wave length dimensions.

## IV. Summary and Discussion

In the present work mean field theories have been applied for analyzing superconductive BEC of excitons in planar systems. I find that Bose condensation in planar or semi planar systems is essentially different from that in bulk media, due to a Bose statistics for the internal momenta of the excitons in the plane, which is different from that for three-dimensional space. Excitons have been described here as electrons-holes pairs producing boson particles and we discuss the effects related to a very large surface density of such excitons. We show that in order to get superconductive exciton BEC for electrons-holes pairs, which will be similar to Cooper-electrons-pairs condensation, a very large number of excitons should be included within a coherent surface with dB wavelength dimensions, and that scattering forces between excitons will be produced.

In Section 2 we started with the general distribution function for boson particles given by (2-5). We notice that for such distribution of excitons in planar systems the chemical potential $\mu$ depends on the surface density of the excitons. I have assumed a simple model of excitons, taken from solid state theory. The equivalent mass of the exciton is given by $m_{eq}$, and $E_{gap} - E_{bind}$ represents a potential barrier between the electron and hole, which are attracted each to the other. Following Bose statistics in the plane, I arrive after some derivations to the distribution (12) which gives by (13) the momentum density $\langle n_k \rangle$ per unit area as function of $m_{eq}$, temperature T and exciton surface density $\langle n \rangle$, where $m_{equ}$ has been defined in (1). An important parameter in the exciton planar distribution is given by $(1-\alpha)^{-1}$ where $\alpha$ has been defined in (14). For very high exciton surface density $\alpha$ becomes very small, and the parameter $(1-\alpha)^{-1}$ is tending to 1. Even under this tendency the effect of $\alpha$ is very important as it eliminates the divergence of the distribution (13) for $k=0$, and its small deviation from 1 is very important parameter in the present analysis.

The main theoretical analysis of Bose condensation of electrons-holes pairs has been given in Section 3. The fundamental result for electrons-holes Cooper pairs Bose condensation has been derived in (41), in which it is shown that super-conductive excitons BEC can be obtained under the condition that this equation can be satisfied. Such result is obtained under the condition that many excitons are included in a surface area with dimensions of dB wavelength and that scattering effects are produced between different excitons included at the coherent area. The transition temperature for Bose condensation is given by (42) by assuming energy gap $\Delta = 0$.

By using the present theoretical model, we might predict superconductivity effects, in electrons-holes bilayer system under certain critical conditions. For such systems one can use the parameter $n = N/A$, given by the surface density of excitons, but exchange the parameter into potential barrier $V_{orth}$, orthogonal to the bilayer plane, preventing electrons-holes rapid recombination. The average potential $V_{parallel} = V_0$ represents attractive interaction in the parallel direction to the bi-layer plane, leading to high surface density of excitons. It seems that excitonic electrons-holes Cooper pairs BEC system with such parameters can be implemented by indirect excitons in traps. Under special conditions analyzed in the present work electrons-holes pairs BEC, with energy gap $\Delta$, can be created, which can lead to superconductivity effects.

**APPENDIX:** The Bose plane integral (given by the integral in (42)), is tabulated as function of the parameters $\left(E_{gap} - E_{bind}\right)/kT$ and $(1-\alpha)^{-1}$. $E_{gap} - E_{bind}$, and $\alpha$ have been defined in (1) and (14), respectively.

| $(1-\alpha)^{-1}$ $\left(E_{gap}-E_{bind}\right)/kT$ | 1.0001 | 1.001 | 1.01 | 1.1 | 4 | 20 |
|---|---|---|---|---|---|---|
| 0.1 | 67.326 | 44.701 | 23,807 | 8.451 | 0.621 | 0.105 |
| 0.3 | 25.373 | 17.754 | 10.422 | 4.299 | 0.370 | 0.063 |
| 1 | 8.368 | 6.072 | 3.815 | 1.768 | 0.177 | 0.031 |
| 3 | 2.938 | 2.172 | 1.412 | 0.697 | 0.077 | 0.0135 |
| 10 | 0.907 | 0.676 | 0.448 | 0.229 | 0.0265 | 0.0047 |

# References

[1] J.M. Blatt, K.W. Boer and W. Brandt, Physical Review **126**, 1691 (1962).

[2] D.W. Snoke, J.P. Wolfe, and A. Mysyrowicz, Phys. Rev. B **41**, 11171 (1990).

[3] L.V. Butov, C.W. Lai, A.L. Ivanov, A.C. Gossard, and D.S. Chemla, Nature **417**, 47 (2002).

[4] J. Kasprzak, M. Richard, S. Kundermann, A. Baas, P. Jeambrun, J.M.J. Keeling, F.M. Marchetti, M.H. Szymanska, R. Andre, J.L. Staehli, V. Savona, P.B. Littlewood, B. Deveaud, and L.S. Dang, Nature **443**, 409 (2006).

[5] D. Snoke, and G.M. Kavoulakis, ArXiv: 1212.4705v1 [cond-mat.quant-gas] Dec 2012.

[6] D.W. Snoke, Hindawi Publishing Corporation, Advances in Condensed Matter Physics, **2011**, 938609 (2011).

[7] D.W. Snoke, Phys. Stat. Sol. **238**, 389 (2003).

[8] S.A. Moskalenko and V. Snoke, *Bose-Einstein Condensation of Excitons and Bi-excitons and Coherent nonlinear optics with excitons* (Cambridge University Press, Cambridge, 2000).

[9] I.E. Perakis, Nature **417**, 33 (2002).

[10] A.A. High, J.R. Leonard, M. Remeika, L.V. Butov, M. Hanson and A.C. Gossard, Nano Letters **12**, 2605 (2012).

[11] A.T. Hammack, N.A. Gippius, S. Yang, G.O. Andreev, L.V. Butov, M. Hanson, and A.C. Gossard, Journal of Applied Physics, **99**, 066104 (2006).

[12] A.A. High, A.K. Thomas, G. Grosso, M. Remeika, A.T. Hammack, A.D. Meyertholen, M.M. Fogler, L.V. Butov, M. Hanson, and A.C. Gossard, Phys. Rev. Lett. **103**, 087403 (2009).

[13] Y.Y. Kuznetsova, A.A. High, and L.V. Butov, Applied Physics Letters **97**, 201106 (2010).

[14] J. Keeling, L.S. Levitov, and P.B. Littlewood, Phys. Rev. Lett. **92** 176402 (2004).

[15] J.P. Eisenstein, and A.H. MacDonald, Nature **432**, 691 (2004).

[16] C. Comte, and P. Nozieres, J. Physique **43**, 1069 (1982).

[17] Y.E. Lozovik, and V.I. Yudson, Sov. Phys. JETP **44**, 389 (1976).

[18] X. Zhu, P.B. Littlewood, M.S. Hybertsen, and T.M. Rice, Phys. Rev. Lett. **74**, 1633 (1995).

[19] P.B. Littlewood, P.R. Eastham, J.M.J. Keeling, F.M. Marchetti, B.D. Simons, and M.H. Szymanska, J. Phys.:

Condens. Matter **16**, S3597 (2004).

[20] Z. W. Gortel, and L. Swierkowski, Surface science **361**, 146 (1996).

[21] Y. E. Lozovik and V. I. Yudson, JETP Lett. **22**, 274 (1975).

[22] S.I. Shevchenko, Surface Science **361**, 150 (1996).

[23] M. Combescot and R. Combescot, Phys. Rev. Lett. **61**, 117 (1988).

[24] C. Schindler and R. Zimmerman, Phys. Rev. B **78**, 045313 (2008).

[25] S. Yang, A. T. Hammack, M. M. Fogler, L. V. Butov, and A. C. Gossard, Phys. Rev. Lett. **97**, 187402 (2006).

[26] A.A. High, J.R. Leonard, A.T. Hammack, M.M. Fogler, L.V. Butov, A.V. Kavokin, K.L. Campman, and A.C. Gossard, Nature, **483,** 584 (2012).

[27] M.M. Fogler, S. Yang, A.T.Hammack, L.V. Butov, and A.C. Gossard, Phys. Rev. B **78**, 035411 (2008).

[28] L.V. Butov, A.C. Gossard and D.S. Chemla, Nature **418**, 751 (2002).

[29] M. Alloing, A. Lemaitre and F. Dubin, EPL **93**, 17007 (2011).

[30] N. Prokofev, and B. Svistunov, Phys. Rev. A **66**, 043608 (2002).

[31] M. Stern, V. Garmider, V. Umansky, and I. Bar-Joseph, Phys. Rev. Lett. **100**, 256402 (2008).

[32] R. M. Lee, N.D. Drummond and R.J. Needs, Phys. Rev. B **79**, 125308 (2009).

[33] R. Maezono, P.L. Rios, T. Ogawa and R.J. Needs, Phys. Rev. Lett. **110**, 216407 (2013).

[34] J. Singh, D. Birkedal, V. G. Lyssenko, and J.M. Hvam, Phys. Rev. B **53**, 15909 (1996).

[35] G. Moody, R. Singh, H. Li, I.A. Akimov, M. Bayer, D. Reuter, A.D. Wieck, A.S. Bracker, D. Gammon, and S.T. Cundiff, Phys. Rev. B **87**, 041304 (2013).

[36] G. Bastard, E. E. Mendez, L. L. Chang, and L. Esaki, Phys. Rev. B. **26**, 1974 (1982).

[37] M. A. M. Versteegh, A.J. van Lange, H.T.C. Stoof and J. I. Dijkhuis, Phys. Rev. B **85**, 195206 (2012).

[38] C. Kittel, *Introduction to solid state physics* (Willey, Hoboken N.J., 2005)

[39] J. C. Kim, and J.P. Wolfe, Phys. Rev. B **57**, 9861 (1998).

[40] H. Ibach, H. Luth, *Solid State Physic* (Springer, Berlin, 1995).